# First results on proton radiography with nuclear emulsion detectors

S. Braccini[a]*, A. Ereditato[a], I. Kreslo[a], U. Moser[a], C. Pistillo[a], S. Studer[a],
P. Scampoli[b,c], A. Coray[d] and E. Pedroni[d]

[a] *Albert Einstein Center for Fundamental Physics,*
*Laboratory for High Energy Physics, University of Bern,*
*Sidlerstrasse 5, CH-3012 Bern, Switzerland*
*E-mail:* `Saverio.Braccini@cern.ch`

[b] *Dipartimento di Scienze Fisiche, Università di Napoli Federico II,*
*Complesso Universitario di Monte S. Angelo, I-80126, Napoli, Italy*

[c] *Department of Radiation Oncology,*
*University of Bern, Inselspital, CH-3010 Bern, Switzerland*

[d] *Paul Scherrer Institut,*
*CH-5232 Villigen PSI, Switzerland*

ABSTRACT: We propose an innovative method for proton radiography based on nuclear emulsion film detectors, a technique in which images are obtained by measuring the position and the residual range of protons passing through the patient's body. For this purpose, nuclear emulsion films interleaved with tissue equivalent absorbers can be used to reconstruct proton tracks with very high accuracy. This is performed through a fully automated scanning procedure employing optical microscopy, routinely used in neutrino physics experiments. Proton radiography can be used in proton therapy to obtain direct information on the average tissue density for treatment planning optimization and to perform imaging with very low dose to the patient. The first prototype of a nuclear emulsion based detector has been conceived, constructed and tested with a therapeutic proton beam. The first promising experimental results have been obtained by imaging simple phantoms.

---

* Corresponding author

# Contents

## 1. Introduction

Proton therapy is becoming a widespread technique in cancer health care, with more than 50 000 patients treated so far in about 30 proton therapy centres in operation worldwide [1]. Due to their characteristic Bragg peak, protons have the feature to come at rest when crossing matter, releasing most of their energy in the last few millimetres of their range. This is the basis of proton therapy, a very precise technique in cancer radiation therapy, which allows to obtain an optimal dose conformation while sparing at best the healthy tissues surrounding the tumour. A crucial issue of this technique resides on the precise knowledge of the proton range. To completely exploit the ballistic properties of protons, an accurate assessment of the location of the Bragg peak is essential for an optimized treatment planning, both in terms of patient positioning relative to the proton beam, and accurate prediction of the dose distribution inside the patient. Presently, X-rays are used for these purposes and, in particular, X-ray Computed Tomography (CT) provides the necessary input information for the calculation of the dose distributions. This implies that the Hounsfield Units (HU), coming from X-ray attenuation, have to be used to determine proton ranges, which, depending on the density of the traversed material, can be estimated only by means of a calibration curve. This procedure introduces uncertainties up to 5% for the proton range (see e.g. [2][3]), thus limiting the exploitation of the peculiar ballistic advantage of charged particles.

The use of high-energy protons to obtain medical images of the patient's body – the so called proton radiography – represents a possible solution to overcome this limitation. The first studies on proton radiography date back to the sixties [4] when it was demonstrated that images of very high contrast could be obtained [5][6] with this technique. The images are based on information related to the average density of the crossed material, allowing a substantial reduction of range uncertainties. Moreover, by means of proton radiography, the dose given to the patient is much less with respect to ordinary X-ray radiography, since every single proton passing through the patient's body brings valuable information [7]. A drawback of proton radiography is its poor spatial resolution due to multiple Coulomb scattering and energy straggling. For this reason and the need of a complex installation to produce high-energy beams, proton radiography was almost abandoned after the pioneering studies in favour of other imaging techniques.

The present widespread of proton therapy has triggered a renewed interest on proton radiography, and new techniques are under study in Europe and USA. A research project was conducted at the Paul Scherrer Institute (PSI) to perform *in vivo* proton radiography with a system consisting of two scintillating fibre hodoscopes and a range telescope made of plastic scintillator tiles [8]. With this system, proton radiography images of an animal patient have been produced [9]. Presently, the Loma Linda University Medical Centre (LLUMC) is collaborating with other institutions to construct and install a detector based on a four-plane silicon track position measuring system, followed by a calorimeter for assessing the proton residual energy. The final goal is to perform a complete Proton Computed Tomography (PCT) [10][11][12][13]. A similar project is underway in Italy [14], based on a silicon micro-strip tracker and on a segmented crystal calorimeter used to characterize particle trajectories and to measure the residual energy, respectively. A different system consisting of a set of position-sensitive Gas Electron Multiplier (GEM) detectors and a scintillator stack read out by solid-state sensors is proposed and developed by the TERA Foundation [15].

In this scientific context and in the SWAN project framework, aiming at the constitution of a combined centre for proton therapy, radioisotope production and research in Bern [16], we propose an innovative technique for proton radiography, which makes use of nuclear emulsion film detectors. The basic principle relies on the irradiation of an Emulsion Cloud Chamber (ECC) detector made of two-sided nuclear emulsions interleaved with tissue equivalent plastic plates and located downstream of the object to be imaged. The method is an end-of-range technique where the position of the Bragg peak is determined by measuring the stop position of the protons inside the detector. This technique exploits the expertise developed for the OPERA neutrino oscillation experiment [17][18]. In this experiment, a renewed use of nuclear emulsion films as a three-dimensional tracking device was made possible thanks to the recent development of industrially produced emulsion films and of fast automated scanning analysis systems [19][20][21].

The proposed technique has a good potential for proton radiography and clinical beam analysis since modern emulsion detectors are rather cheap and easy to handle. These characteristics are surely valuable in a clinical environment where daily routine treatments have absolute priority. A simple and useful application is the characterization of proton beams in the commissioning phase of proton therapy clinical facilities, when the purchase and operation of expensive and sophisticated detectors could not be justified for only this purpose. A clear limitation of emulsion based detectors is the absence of a real-time response. However, the speed of modern scanning systems is steadily increasing and good prospects are expected for the future [22].

In this paper we report for the first time on the design, construction and first experimental results on proton radiography obtained with an emulsion detector prototype.

## 2. Nuclear emulsions and their potential for proton radiography

Nuclear emulsions have a very long track record as particle detectors [23] and represented an important tool for fundamental discoveries. They consist of a gel with silver bromide (AgBr) crystals where a latent track is formed after the passage of an ionising particle. Tracks can be visualised and analysed by optical microscopes after a specific chemical development process performed under controlled conditions. The development allows the formation of a sequence of silver grains along the particle track that, with dimensions of 1 μm or less, are well visible by means of optical microscopes.

Due to their micrometric space resolution, nuclear emulsions still represent the most precise particle tracking device. Even within a single 50 μm layer, they allow full three-dimensional reconstruction of particle tracks and the measurement of energy loss. This is performed by evaluating the silver grain density along the particle track.

In the past, the main limiting factor in the use of these detectors was the time consuming manual scanning procedure. The recent impressive development of fast automated scanning systems [19][20][21] has fostered a renewed interest in this technique, especially in neutrino physics, where the OPERA experiment [18][24] is aiming at the first direct observation of neutrino oscillation by means of nuclear emulsion film detectors. In the recent years, nuclear emulsions have been successfully applied in several other research fields [22] and in particular to hadrontherapy to investigate nuclear fragmentation of therapeutical carbon ion beams [25][26].

On the basis of the high performing tracking features of nuclear emulsion films and on the presently available automatic scanning systems, the application of this technique to proton radiography represents an interesting possibility. The basic principle is presented in Fig. 1, where an ECC made of a modular structure of emulsion films interleaved with tissue equivalent absorbers is exposed to a mono-energetic proton beam passing through the object to be imaged.

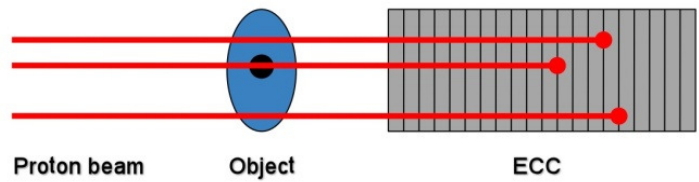

Fig. 1 - *Principle of proton radiography with nuclear emulsion film detectors. The range of the individual protons is measured by the number of crossed emulsions in the detector. Higher traversed material densities correspond to shorter ranges.*

The emulsion detector has to be thick enough to contain the end-of-range points of all the protons. The difference in density encountered by protons translates into a difference on the detected range. Once exposed, developed and automatically scanned, the emulsions allow obtaining digital data from which proton range measurements can be performed and used to build a proton radiography image.

## 3. Construction and test of the first prototype detector

Two phantoms to be imaged have been constructed in order to perform a proof of principle experiment in a simple geometry. The first phantom (Fig. 2, left) is made of polymethylmethacrylate (PMMA) and presents a simple 1 cm step ("step phantom"). In this way, protons traverse only 3 or 4 cm of PMMA. The second phantom (Fig. 2, right) has a total thickness of 4.5 cm and contains five 5 × 5 mm$^2$ aluminium rods positioned at different depths in a PMMA structure ("rod phantom").

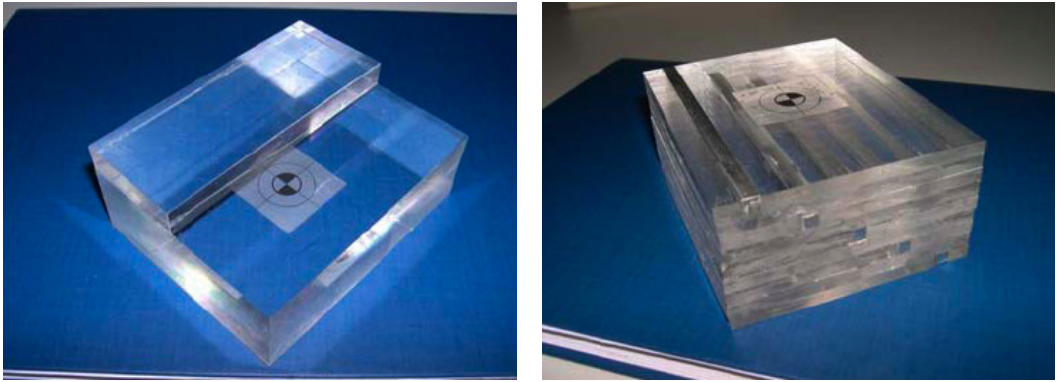

Fig. 2 - *The "step" (left) and the "rod" (right) phantoms used to obtain the first proton radiography images.*

The emulsion films used for this study consist of 2 layers of 44 μm thick sensitive emulsion coated on both sides of a 205 μm thick triacetate base with a surface of $10.2 \times 12.5$ cm$^2$. As tissue equivalent absorber material, polystyrene plates with the same transverse dimensions as the emulsions and a thickness of 0.54 mm are used.

To check the feasibility of this new application of nuclear emulsion films, simulations based on the GEANT3 toolkit [27] have been performed. Several configurations of ECCs made of nuclear emulsions and polystyrene layers have been considered in order to optimize the resolution on the proton range and to minimize the number of emulsions to be employed [28].

A proton beam with energy of 138 MeV - as the one used for the beam tests - and with a flat distribution over the $10.2 \times 12.5$ cm$^2$ sensitive surface has been simulated.

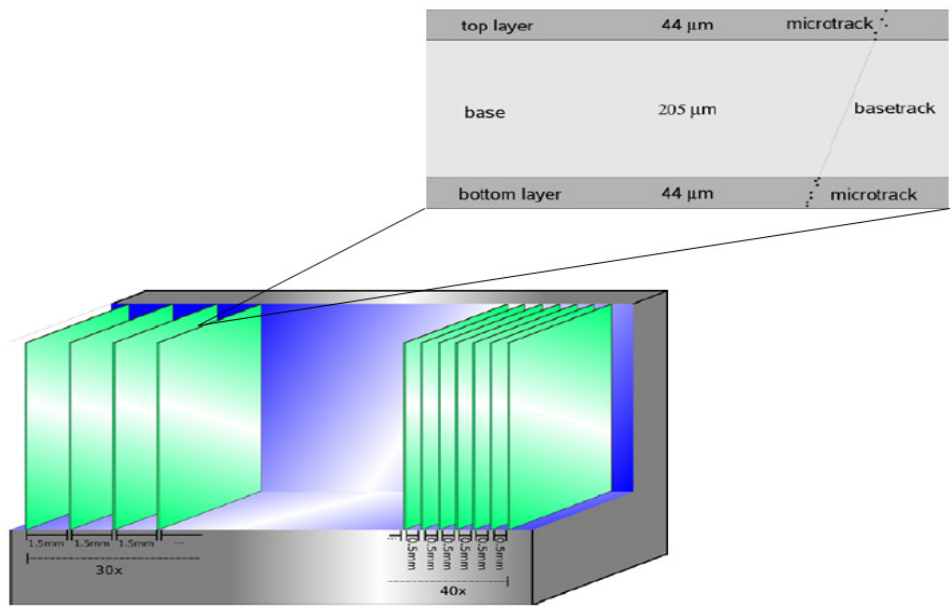

Fig. 3 – *Detector structure: Emulsion film plates are interleaved with blocks of absorber (not shown) of different thickness. The structure of one emulsion film is highlighted in the inset together with the two micro-tracks produced in the two sensitive emulsion layers.*

The results of the simulations led to the design of the detector to be coupled with the two phantoms. As presented in Fig. 3, the apparatus is divided into two parts. The first one is made of 30 emulsion films, each interleaved with 3 polystyrene absorber plates, allowing tracking of passing through protons using a minimal number of films. The second part is composed of 40 films, each interleaved with one polystyrene plate, allowing a precise measurement of the proton range in correspondence of the Bragg peak. The overall dimensions of the detector are $10.2 \times 12.5$ cm$^2$ in the transverse direction, and 90.3 mm along the beam. As shown in Fig. 3, micro-tracks are 3D segments reconstructed in each sensitive layer of an emulsion plate while the base-track is the line connecting the two points of the micro-tracks closest to the plastic base. This allows to get rid of possible distortions [18].

Simulation results for the system composed of the emulsion detector and the "step phantom" are reported in Fig. 4. All proton tracks stop in the second part of the detector, where spatial resolution is maximal. It has to be noted that the phantom cross section is smaller than the surface of one emulsion. In this way protons on the borders cross the entire detector and provide the necessary information for the relative alignment of the emulsions.

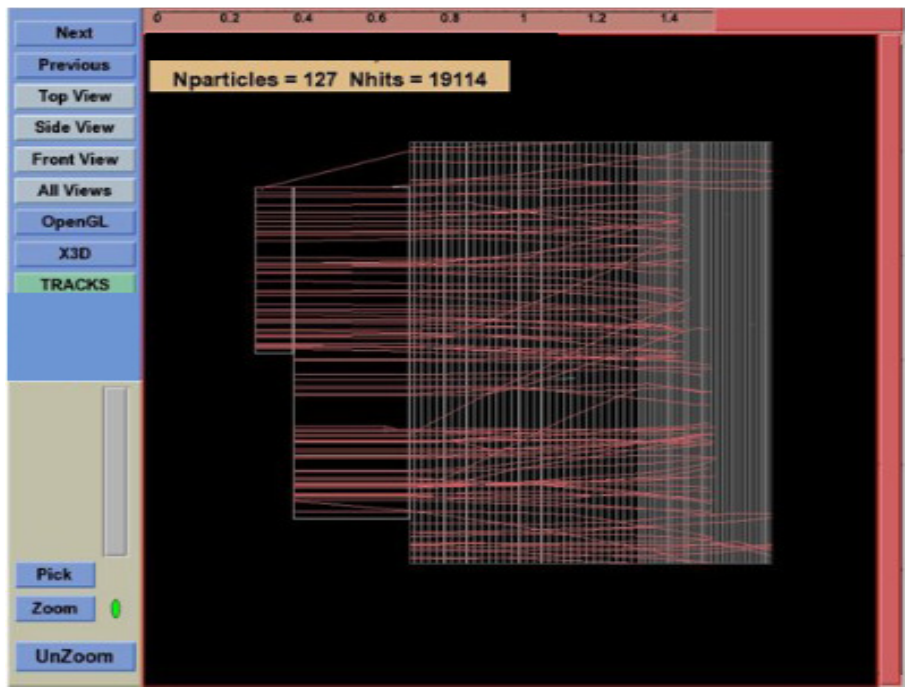

Fig. 4 - *Simulated proton tracks in the detector behind the "step phantom" clearly show the difference in range.*

For the evaluation of the proton range in the detector, a simulation has been performed using SRIM [29], a software package for the calculation of stopping power and range of ions in matter. For the "step phantom" the ranges inside the detector are found to be 76 mm and 86 mm for protons crossing 4 cm and 3 cm of PMMA in the phantom, respectively. For the "rod phantom" the calculated ranges are 68 mm and 72 mm for protons crossing one aluminium rod or PMMA only, respectively.

The assembly of the two detectors has been performed in a dark room by piling up the emulsion films and the absorbers on a flat marble table. To enhance rigidity, two 1 mm thick aluminium plates have been located at the top and at the bottom of the stack. After the verification of the alignment of emulsions and absorbers to be within 0.5 mm, the detectors have been closed using a special plastic foil, black towards the emulsions and reflective externally. In this way, the detectors are light tight. The two detectors have been stored in a laboratory located at LHEP 40 m underground before irradiation with the proton beam to minimize cosmic ray background.

## 4. Experimental results

The clinical beam of the Gantry 1 [30] at PSI has been used for the first beam tests, as shown in Fig. 5. In order to be able to reconstruct proton tracks without saturating the emulsions, the number of protons has to be limited to $10^6$ protons on a $10 \times 10$ cm$^2$ surface. This corresponds to an average dose to the patient of the order of $10^{-5}$ Gy. For this purpose a special low intensity beam setting has been used and an accurate measurement of the number of protons per unit area performed with a coincidence scintillating telescope.

The two phantom-detector systems have been exposed to 138 MeV protons uniformly spread on the surface by means of spot scanning. The gantry was adjusted to deliver the beam perpendicular to the emulsions.

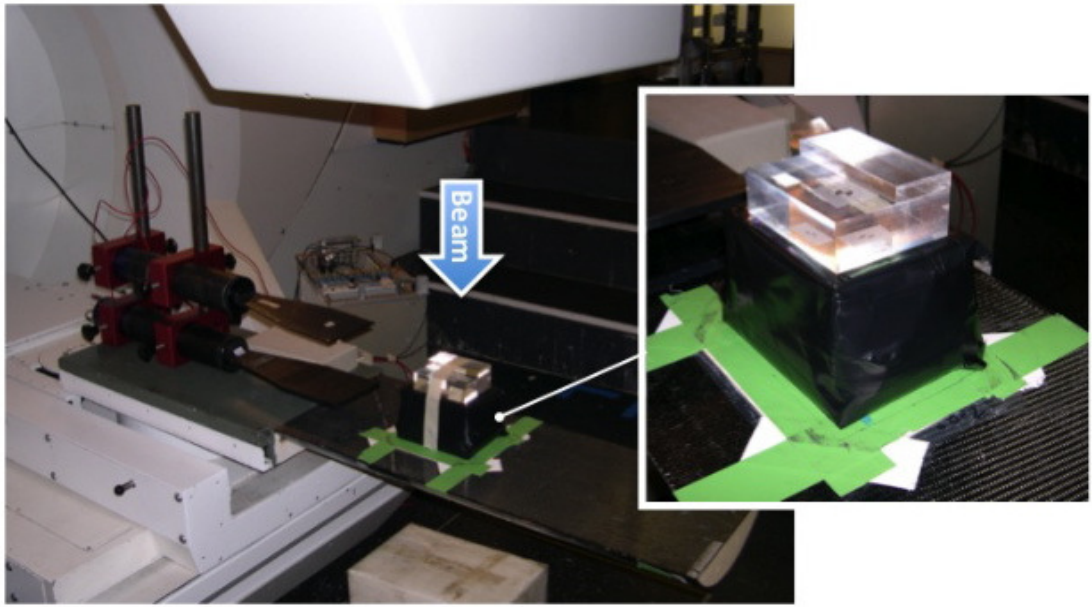

Fig. 5 - *First beam tests of proton radiography with nuclear emulsions at the Gantry 1 at PSI. The detector with the "step phantom" on the top is highlighted in the inset. The scintillator telescope used to verify the number of incoming protons is visible on the left.*

After the irradiation, the two detectors have been brought back to LHEP where they were disassembled and the emulsions misaligned in order to minimize the background due to cosmic-rays. In this way, cosmic muons do not produce reconstructable tracks. The emulsions have then been put in closed containers and transported to the Gran Sasso National Laboratory (LNGS) of the Italian Institute for Nuclear Physics (INFN) where they have been developed following the same procedure used for the OPERA experiment [18].

The emulsion analysis has been performed by means of automatic scanning systems at LHEP. Here a specifically conceived laboratory equipped with five European Scanning System (ESS) [20][21][31] microscopes is active for routine scanning of the emulsions of the OPERA experiment. These optical microscopes are equipped with dry objectives [32], computer driven motorized stages, CMOS cameras to grab images along the depth of the emulsion layer, and automatic robots for handling the emulsion films [33]. The equipment developed at LHEP allows to reach scanning speeds of 10 cm$^2$ per hour with prospects for a factor 10 improvement in the next years [22]. Silver grains due to ionization are recognized along the path where charged particles crossed the emulsion. Sequences of grains are then identified and track information obtained and stored. The FEDRA (Framework for Emulsion Data Reconstruction and Analysis) program [34] has been used for the off-line reconstruction of proton trajectories inside the emulsion plates.

At the energies used for this measurement, proton ionization is significantly higher than the one of cosmic-rays. As presented in Fig. 6, the number of grains along the track, together with the angular factor of merit, allows the identification of protons over a negligible background. The angular factor of merit is a parameter based on the angular matching between the two micro-tracks inside one emulsion plate.

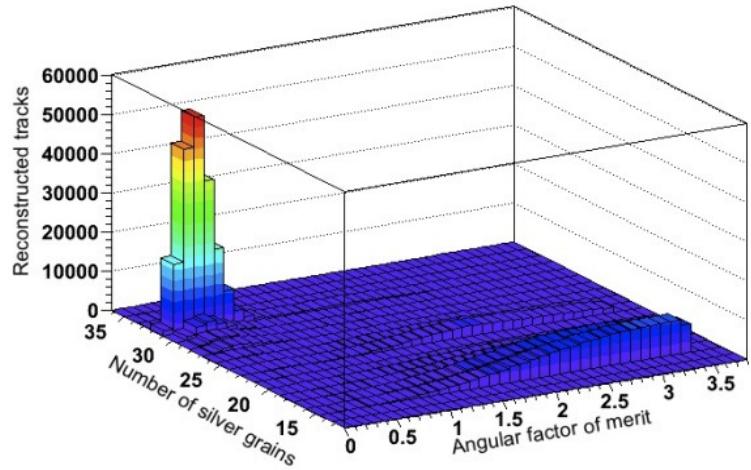

*Fig. 6 – The number of silver grains and the angular factor of merit allow the selection of the proton tracks over a negligible background.*

The proton density has been measured to be about 7 × 10$^3$ protons/cm$^2$, in good agreement with the result obtained with the scintillating telescope before the irradiation.

For imaging the "step phantom", an overall area of 25 × 30 mm$^2$ has been scanned for each emulsion and all the proton tracks identified. The detector has been analysed by subdividing it in columns along the beam direction. The transverse area of each column is 1 mm$^2$.

The variation of the proton density as a function of the depth has been evaluated for each column. To calculate the residual range, data have been fitted for each column with the following Fermi-Dirac function:

$$f(z) = A - \frac{A}{e^{\frac{R-z}{S}} - 1}$$

where z is the coordinate along the beam, A, R and S are free parameters. In particular, R represents the proton residual range. An example of one fitted distribution is reported in Fig. 7.

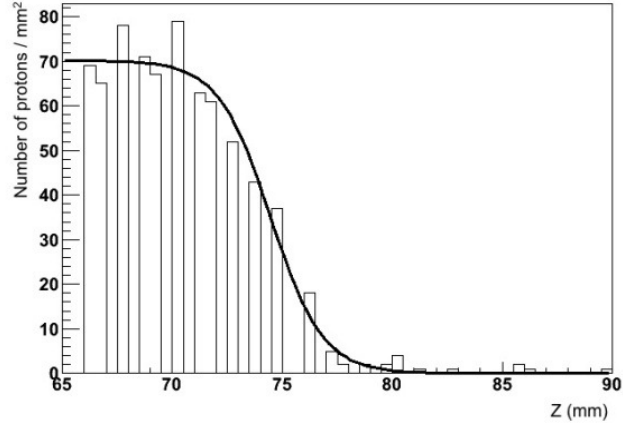

Fig. 7 – *The number of identified protons as a function of the z coordinate inside a column of 1 mm$^2$. The fit allows the localization of the Bragg peak.*

Once the residual range is obtained for all the columns, an image of the phantom can be obtained by plotting R as a function of the transverse coordinates, as presented in Fig. 8. The shape of the phantom is well reproduced, and the residual ranges inside the detector in correspondence of the two thicknesses of the "step phantom" are measured to be (75.0±0.6) mm and (84.8±0.8) mm, where errors have been evaluated by the fitting procedure. These results are in good agreement with the values obtained with the Monte Carlo simulations, as discussed in Section 3.

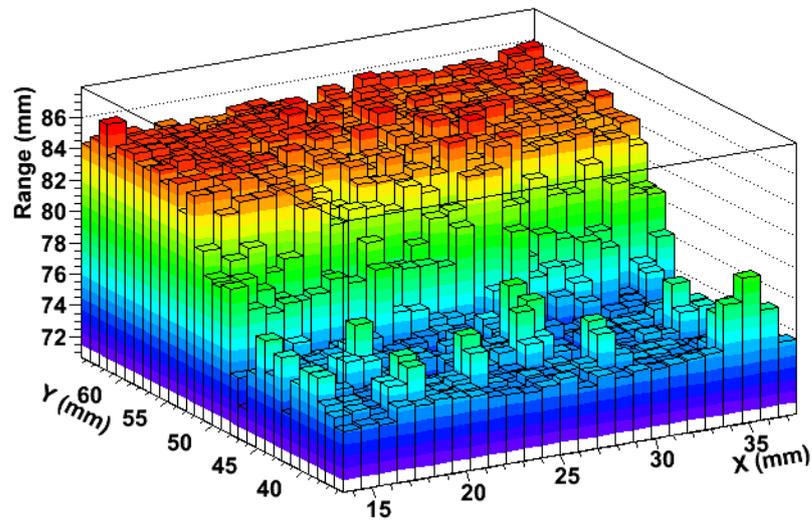

Fig. 8 - *The measured proton range as a function of the transverse coordinates shows a clear proton radiography image of the "step phantom".*

The same approach is used for the analysis of the "rod phantom". In this case, an area of 30 × 75 mm$^2$ has been scanned for each emulsion film and columns of 1 mm$^2$ transverse dimension have been defined, as for the "step phantom". After fitting, a clear image of the "rod phantom" is obtained, as presented in Fig. 9. The residual range in the detector for protons crossing a rod is measured to be (69.6±1.2) mm and (71.9±0.6) mm for those crossing PMMA only, in good agreement with the simulations.

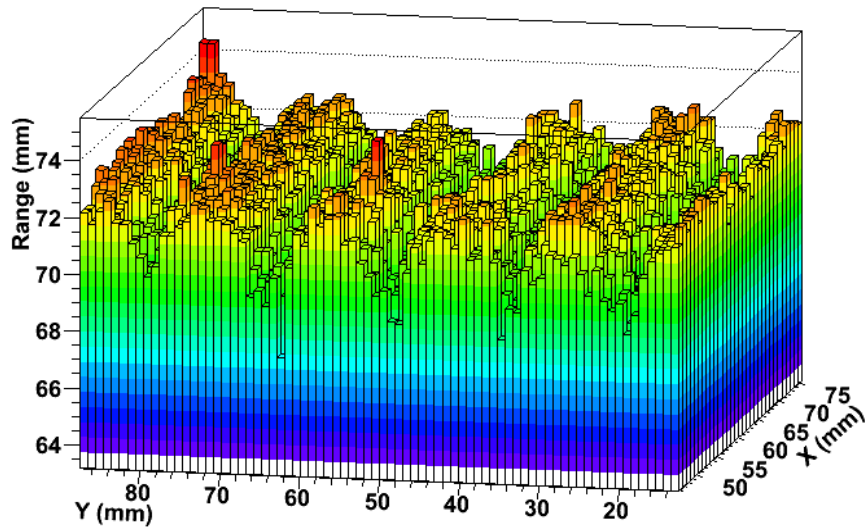

Fig. 9 - *The measured proton range as a function of the transverse coordinates give a proton radiography image of the "rod phantom".*

## 5. Conclusions

We propose an innovative method based on nuclear emulsion film detectors for proton radiography. Prototype detectors made of nuclear emulsion films interleaved with tissue equivalent absorbers have been conceived, constructed and successfully tested with the clinical proton beam of the Gantry 1 at PSI.

For the first time, clear images of two phantoms have been obtained, proving the principle of this technology and opening the way to more sophisticated tests and applications that will be the subject of forthcoming studies.

**Acknowledgments**

The authors would like to acknowledge the LHEP technical staff and the emulsion development team of the OPERA experiment.

**References**

[1] U. Amaldi et al., *Accelerators for hadrontherapy: from Lawrence cyclotrons to linacs*, accepted for publication by Nucl. Insrt. Meth. A (2010).

[2] U. Schneider and E. Pedroni, *Proton radiography as a tool for quality control in proton therapy*, Med. Phys. (1995) **22** 353.

[3] B. Schaffner and E. Pedroni, *The precision of proton range calculations in proton radiotherapy treatment planning: experimental verification of the relation between CT-HU and proton stopping power, Phys. Med. Biol.* (1998) **43** 1579.

[4] A. M. Koehler, *Proton Radiography, Science* (1968) **160** 303.

[5] V.W. Steward and A.M. Koehler, *Proton Beam Radiography in Tumor Detection, Science* (1973) **179** 913.

[6] J. A. Cookson, *Radiography with protons, Naturwissenschaften* (1974) **61** 184.

[7] K. M. Hanson et al., *Computed tomography using proton energy loss, Phys. Med. Biol.* (1981) **26** 965.

[8] P. Pemler et al., *A detector system for proton radiography on the gantry of the Paul-Scherrer-Institute, Nucl. Instr. Meth A.* (1999) **432** 483.

[9] U. Schneider et al., *First proton radiography of an animal patient Med. Phys.* (2004) **31** 1046.

[10] H. F. -W. Sadrozinski et al*., Issues in Proton Computed Tomography, Nucl. Insrt. Meth. A* (2003) **511** 275.

[11] L. Johnson et al., *Initial studies on proton computed tomography using a silicon strip detector telescope, Nucl. Instr. Meth. A* (2003) **514** 215.

[12] R. Schulte et al., *Conceptual Design of a Proton Computed Tomography System for Applications in Proton Radiation Therapy, IEEE Trans. Nucl. Sci*. (2004) **51** 866.

[13] H. F. -W. Sadrozinski et al*., Toward Proton Computed Tomography, IEEE Trans. Nucl. Sci.* (2004) **51** 3.

[14] C. Talamonti et al., *Proton radiography for clinical applications, Nucl. Instr. Meth.* (2010) **A612** 571.

[15] U. Amaldi et al., *Advanced Quality Assurance for CNAO,* (2009) *Nucl. Instr. and Meth. A* doi:10.1016/j.nima.2009.06.087.

[16] S. Braccini, D. M. Aebersold, A. Ereditato, P. Scampoli and K. von Bremen, *SWAN: a combined centre for radioisotope production, proton therapy and research in Bern, presented at the Workshop on Physics for Health in Europe, CERN, Geneva, February 2010.*

[17] A. Ereditato, K. Niwa and P. Strolin, *OPERA: an emulsion detector for a long baseline nu(mu)-nu(tau) oscillation search, Nucl. Physics B Proceedings supplement* (1998) **66** 423.

[18] R. Acquafredda et al., *The OPERA experiment in the CERN to Gran Sasso neutrino beam*, JINST (2009) **4** P04018.

[19] S. Aoki et al., *Fully automate emulsion analysis system, Nucl. Instr. Meth. B* (1990) 51 466.

[20] N. Armenise et al., *High speed particle tracking in nuclear emulsion by last-generation automatic microscopes, Nucl. Instr. and Meth. A* (2005) **551** 261.

[21] L. Arrabito et al., *Hardware performance of a scanning system for high speed analysis of nuclear emulsions, Nucl. Instr. and Meth. A* (2006) **568** 578.

[22] G. De Lellis, A. Ereditato and K. Niwa, Nuclear Emulsions, *to appear in The Handbook of Particle Physics, Vol. II, Springer Verlag (Landolt-Boernstein), Heidelberg – Berlin, Editor H. Schopper.*

[23] F. Powell, P. H. Fowler, and D. H. Perkins*, The Study of Elementary Particles by the photographic Method, Pergamon Press, New York, 1959.*

[24] T. Nakamura*, The OPERA film: New nuclear emulsion for large-scale, high-precision experiments, Nucl. Instr. and Meth. A* (2006) **556** 80.

[25] T. Toshito et al., *Measurements of total and partial charge-changing cross sections for 200- to 400 MeV/nucleon $^{12}$C on water and polycarbonate*, Phys. Rev. C (2007) 75 054606.

[26] G. de Lellis et al., *Emulsion Cloud Chamber technique to measure the fragmentation of a high-energy carbon beam*, JINST 2 (2007) P06004.

[27] *GEANT3 Detector description and simulation tool*, http://wwwasdoc.web.cern.ch/wwwasdoc/geant_html3/geantall.html (1995)

[28] S. Studer, *Study of a nuclear emulsion based detector for proton-radiography*, Master Thesis, University of Bern, 2009.

[29] J. F. Ziegler and J. M. Manoyan, *The stopping of ions in compounds*, Nucl. Instr. and Meth. B (1988) **35** 215.

[30] E. Pedroni et al., *The 200-MeV proton therapy project at the Paul Scherrer Institute: conceptual design and practical realization*, Med. Phys. (1995) **22** 37.

[31] L. Arrabito et al., *Track reconstruction in the emulsion-lead target of the OPERA experiment using the ESS microscope*. JINST 2:P05004 (2007).

[32] I. Kreslo et al., *High-speed analysis of nuclear emulsion films with the use of dry objective lenses*. JINST 3:P04006 (2008).

[33] K. Borer at al., *A novel automatic film changer for high-speed analysis of nuclear emulsions*, Nucl. Instrum. Meth. A (2006) **566** 327.

[34] V. Tioukov et al., *The FEDRA - Framework for Emulsion Data Reconstruction and Analysis in the OPERA experiment*. Nucl. Instr. Meth. (2006) **A559** 103.